# The Diffusion of Sb into Ge without Contamination by Fast Diffusing Electrically Active Impurities.

D. Alexiev[1], D. A. Prokopovich[1,2] and L. Mo[1]

[1]Australian Nuclear Science and Technology Organisation (ANSTO), PMB 1 Menai NSW 2234 Australia
[2]University of Wollongong, Wollongong, N.S.W. 2522 Australia

## ABSTRACT

A method has been developed to permit the diffusion of Sb into Ge at high temperatures (~850 °C) without contamination by fast diffusing electrically active impurities in particular by Cu. A liquid metal alloy is used as a getter of Cu and other fast diffusing impurities. This alloy, Ga- In eutectic, completely encloses the Ge sample although in physical contact on only one face.
The behaviour of Cu as a contaminant in Ge and the methods known to prevent and extract (or gather) Cu contamination are reviewed briefly. Preliminary experiments are described which demonstrate the difficulty of removing fast diffusing impurities in spite of the use of liquid metal getter (Ge-In and Au). The advantages and disadvantages of the technique are discussed.

## 1. Introduction

The information obtained from the experiments described later in this paper is in itself insufficient to allow the absolute identification of the contaminating impurity or impurities in the samples. Rather, the impurity or impurities are characterised by a high diffusion coefficient and a high solubility in certain liquid metal. At room temperature it is an acceptor. These characteristics, however, together with data presented in the literature enable us to identify the likely impurity as Cu.
Those properties of Cu in Ge which make it a serious contaminant are presented below.

**Properties of Cu in Ge**
Cu in Ge is electrically active. The probability of ionisation is temperature dependant. It can exist in two forms; as an acceptor when the atom is substitutional, or as a donor when it is interstitial. The acceptor level is triply charged while the donor is singly charged.
We can summarise the information contained in the proceeding paragraphs by saying that the high diffusion coefficient of Cu in Ge enables it to penetrate deep into the bulk of samples raised to high temperatures (> 600 °C). At room temperature the sample will show acceptor contamination and a reduced lifetime.

**Diffusivity of Cu in Ge**
Cu exhibits an unusually large diffusion coefficient. This is due to the rapid diffusion of the interstitial atom. However, since an interstitial atom may convert to a substitutional atom at a vacancy, the diffusion may depend on the vacancy concentration of the starting material. Also, the diffusion of Cu may depend on the concentration of other electrically active dopants (for pure but still extrinsic Ge eg. less than $10^{16}$ impurity atoms $cm^{-3}$). The diffusivity for interstitial Cu is $4 \times 10^{-3}$ $cm^2$ $s^{-1}$ at 850 °C, while the activation energy is 0.33eV. Substitutional Cu is relatively immobile and its diffusion can be neglected.

**Contamination of Ge by Cu**
The significance of Cu as an electrically contaminating impurity of Ge was realised by the beginning of last decade. At the present time, although the problem is much better understood, techniques are such that in certain applications it is still difficult to reduce Cu contamination to an insignificant amount. These topics are discussed below.

While it is quite possible to purify Ge so that it is intrinsic at room temperature, the majority of semiconductor devices do not require material of this purity. In such devices Cu contamination is generally not too serious a problem. High resolution radiation detectors however, ideally require material of very high purity. In Ge detectors for example the only doping agents should be intentional, and with shallow ionisation levels, otherwise the resolution is degraded. Compensating dopants, such as Cu and Li have been shown [1] to be significant contaminating impurities in the growth of high purity single crystal of Ge. Furthermore, contamination from Group 1 ( Ag, Au, Cu )  and Group IV ( Si, Ge, Sn ) during high temperature annealing of II-VI semiconductor materials will degrade significantly detector performance [ 2 ]. Experiments to determine the properties of thermally induced vacancies in Ge are difficult to perform because of Cu contamination. Such experiments generally involve raising the temperature of a sample to a high value in order to obtain a high concentration of vacancies.  Quenching to a low temperature produces a super saturated solution of vacancies in Ge. The kinetics of the vacancies are than studied. Vacancies have acceptor properties but can be annihilated in the  reaction
$$Cu^i + V \longrightarrow Cu^s$$
Diffusions of conventional dopants into Ge ( to form contacts on radiation detectors, for example) must be performed at high temperatures. Such dopants include Ga, In, B, As, Sb and P. These intentional shallow dopants all have small diffusion coefficients ($5 \times 10^{-9}$ – $5 \times 10^{-12}$ $cm^2$ $s^{-1}$ at 850° C which is $10^6$ – $10^9$ below that of Cu).

In experiments where pure Ge samples undergo excursions to high temperatures precautions must therefore be taken to reduce the chance of Cu contamination. Technique are available to getter Cu from contaminated material. The prevention of Cu contamination and the extraction of Cu from contaminated material are discussed in the following sections.

**Prevention of Cu Contamination**
Several procedures are available to prevent or at least reduce Cu contamination when a high temperature cycle is necessary. More than one of these procedures are normally used in a attempt to further reduces the contamination. These procedures are listed below.

Of paramount importance is the surface cleanliness and purity of the components of the system. The various components of the system  which will be raised to a high temperature should be chosen with the minimum concentration of Cu and any other fast diffusing electrically active impurities. Care should be taken during the preparation of the components so that the surface is as clean as possible. Wherever possible, material should be etch-cleaned using high purity acids and pure (distilled or deionised ) water. It may be possible heat clean some components. This should be done in an oxygen atmosphere or an oxidising furnace. Such treatment will permit the oxidation of Cu.

An important method of obtaining surface cleanliness involves the use of aqueous KCN. Cu forms salts which are soluble in water. It is thus possible to reduce surface contamination of the sample and other components. Furthermore, a solution of KCN has been shown to remove Cu impurity of submersed Ga in liquid form [3]. Experiments by other researchers showed that heating samples to an intermediate temperature in an oxygen atmosphere with a trace of water vapour, initially increased the lifetime of the sample. Assuming that this process getters Cu from Ge other experiments used this surface oxidation technique to prevent and getters Cu contamination .

Certain metals which are liquid at high temperatures can be used to prevent Cu diffusing into Ge. These are discussed in the next section on extraction.

Similarly certain metals which are still in a solid phase will prevent Cu diffusing into Ge. These are also discussed in the next section.

**Gettering or Extraction of Cu contamination.**

Ge samples known or suspected to be contaminated may be gettered by the following processes. These gettering processes are all aided by the high diffusion coefficient of Cu in Ge.

A prolonged period at a intermediate temperature will allow Cu to diffuse out from the bulk of the sample. If the temperature of the surrounds are lower than that of the sample Cu can condense and so be effectively removed. When using this method of gettering it is important that there is no additional source of Cu in the system.

At temperatures below the Ge melting point certain metals may be liquid. Or else the eutectic temperature of the Ge metal system may be below the operating temperature. If either of these two conditions hold there exists a a liquid metal-rich region in contact with the semiconductor. If the solubility of Cu in the metal or in the liquid alloy with Ge is very much greater then that of Cu in Ge such liquid metals will getter Cu from Ge. The diffusion coefficient of such metals must be much smaller than that of Cu. The solubility of Ge in the liquid phase should be small so as to minimise loss of Ge by alloying. The distribution coefficient for Cu in the Cu-Ge metal ternary system should be small to allow accumulation of Cu in the liquid. Examples of such gettering agents are Au, Sn, Pb, Ga and In.

There have been reports on the gettering of Cu from Ge with metals which are in the solid phase. The solubility of Cu in these metals must be greater than that in Ge. Examples are Fe and Rh.

## 2. Preliminary Experiments

A program of work was begun in an effort to produce stable, robust, non-injecting n+ contacts for Ge ray γ-detectors. High temperature diffusions of Sb were attempted in a system shown in Figure 1. The underside of the Ge sample was normally painted with Ga- In eutectic which is liquid at room temperature. These liquid metals were intended to getter Cu from the sample since the diffusion coefficient of both Ga and In is smaller by ~$10^2$ than that Sb. It should have been possible to completely coat the sample with the liquid and produce a diffused contact. However the depth of the Ga-In-Ge alloy prevented the complete coating of the sample.

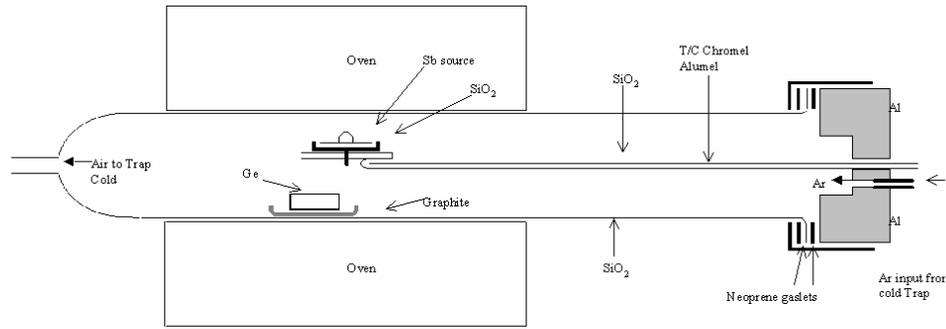

Figure 1: System used to diffuse Sb into Ge

Diffusions were carried out at ~850 °C for one or two hours giving diffusion depth of 40 or 50 μ. However, examination of diffused samples showed an acceptor gradient from the open or Sb- diffused face to the gettered face. This fact was shown by measurements of resistivity ( with 4 pt. Probe) and majority carrier type ( with the thermal probe) . These measurements were made at room temperature and enabled a qualitative estimate of the doping impurity concentration to be made. The profile of the impurity from open to gettered face was examined in part by lapping and probing repeatedly. Figure 2 typically shows results obtained underneath the two faces for some distance into a sample.

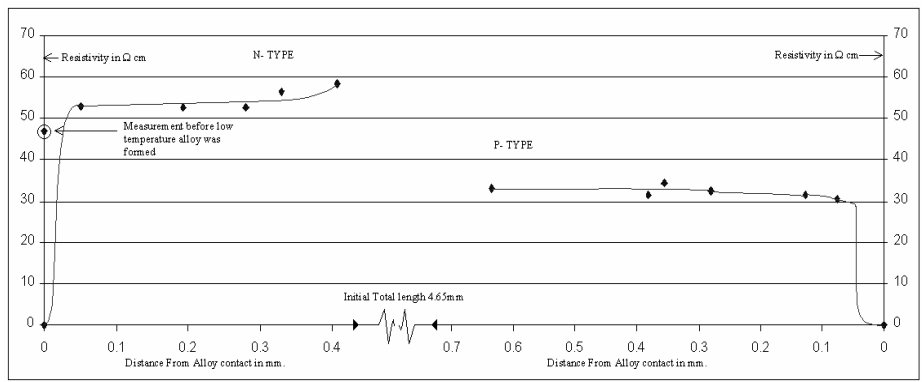

Figure 2: Resistivity measurement of Sb in N(111).

With these results in mind a series of experiments were undertaken in an attempt to locate the Cu source, or sources, and to prevent the contamination by Cu. Considerable care was taken to ensure that the components of the system were of high purity and that their surfaces were clean. All glass-ware was of high purity quartz. This was etch-cleaned with 4:1 / $HNO_3$ : HF . Deionised or doubly- distilled water was used for diluting the tech and for final washing.
The following were suspected of being Cu sources;
The $SiO_2$ furnace tube and thermocouple holder, the thermocouple which was chromel alumel, and the carbon plates on which the sample and Sb source sat. The carbon was electronic grade graphite.
The possibility of each of these as sources of Cu was examined in the following experiments. The carbon plate underneath the sample was replaced with a clean $Al_2O_3$ plate. This plate was cleaned in 4:1 HNO$_3$ : HF enchant and washed with doubly distilled water. It was than heat cleaned with an oxygen rich flame.

The furnace was calibrated and the thermocouple and it's SiO$_2$ holder, the second carbon plate, and SiO$_2$ dish were removed. The Sb source was not required in these Cu-source-locating experiment.

A tube of Al$_2$O$_3$ was etch cleaned, washed and inserted inside the SiO$_2$ furnace-tube. The furnace was brought up to ~850° C in a preliminary cycle in order to heat clean this tube. Oxygen was passed through the furnace in order to oxidise any Cu present in the Al$_2$O$_3$. In addition it was hoped that oxygen would be trapped in the Al$_2$O$_3$ and the during the actual experiment run any free copper in the system would be removed by oxidation.

These experiments were performed in the order described and the modifications in an experiment included those of the proceeding experiment. The standard practice was to etch-clean in 4:1 / HNO$_3$ : HF and wash in doubly ionised H$_2$O all components before an experiment was performed.

These experiments while affecting the amount of Cu slightly did not point to any particular source. A gradient of Cu was always present as the samples were formed to be p-type underneath the span face and n-type under the alloyed face. There was some evidence of out diffusion as under the open face the material decreased in p-type resistivity with depth.

During these experiments it was confirmed that Au gettered Cu. However since the diffusion coefficient of Au is a factor of 30 greater than that of Sb it is not possible to completely cover the sample with Au and diffuse an Sb contact. The Au was electroplated onto the sample.

## 3. Complete Prevention of Cu contamination

A method was finally found which allowed an Sb diffusion without contamination by Cu. The principle was to totally enclose the Ge sample in a liquid metal getter. This was done by painting with liquid Ga-In eutectic the inside of an Al$_2$O$_3$ 'pill-box'. The sample painted on one face, was placed inside this pill-box ( painted face downwards).

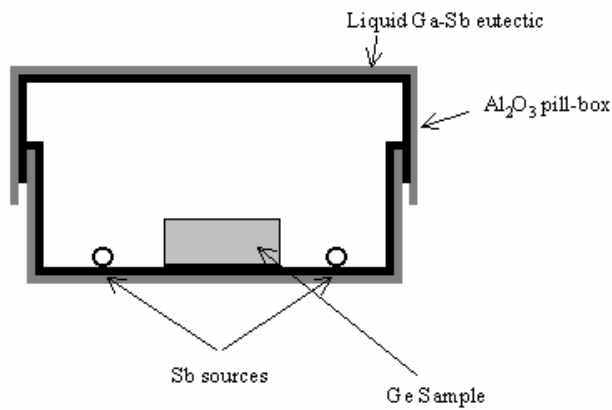

Figure 3: Al$_2$O$_3$ pillbox.

Figure 3 indicated the system used. Small Sb sources were laid on top of the eutectic layer on the floor of the box. The pill box was placed in the centre of the furnace described in Figure 1 and the temperature brought up to ~850 °C. The graphite plates and quartz dish were unnecessary and therefore were removed. Since the furnace had been calibrated the thermocouple and it's SiO$_2$ holder were withdrawn.
Diffusions were carried out for one or two hours at 850 °C. the sample was brought up to 850 °C in two hours allowed to cool naturally. The pill-box and sample were removed from the furnace when the temperature was below 100 °C (some 4 hours later). Results such as those shown in Figure 4 were obtained for several sample. As previously, resistivity and thermal probe measurements were made after each of several laps of the material under each face. In this way the doping profile was examined. No change in polarity or resistivity was formed.

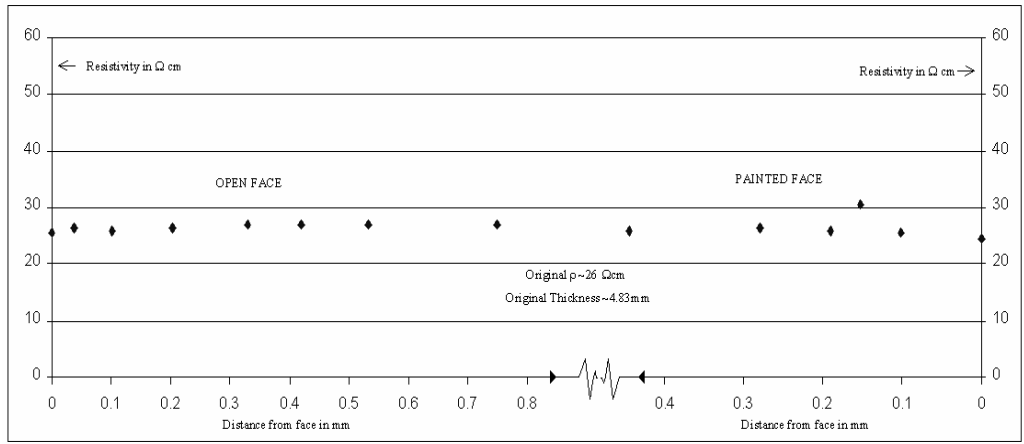

Figure 4: Resistivity measurement of n-type sample. Al$_2$O$_3$ Pb painted internally with Ga-In eutectic

The successful gettering action of this method was confirmed in two ways. Firstly a sample from one of the preliminary experiments, having a contamination gradient from one to the other was inserted in the pill box in the manner described above. The Sb sources were omitted . the original resistivity and polarity were regained after 2 hours at 850 °C inside the Ga-in painted pill-box. The results are shown in Figure 5. This experiment was repeated successfully with a second contaminated sample.
In another test of the gettering ability of the Ga-In as used in this method, samples were electro-plated with Cu and then raised to 850 °C for 2 hours in the furnace. This allowed uniform Cu contamination by diffusion . These samples were then subjected to the treatment previously described. The original resistivity was again regained.

## 4. Discussion and Conclusion

This method permits the diffusion of Sb into Ge. The degree of contamination is small but it may not be zero. The unwanted acceptor density is certainly less than the intentional shallow dopant density since the high temperature cycle can be carried out without a change in resistivity. Measurements of minority carrier lifetime would permit a more sensitive examination of the impurity content. In addition, examination of the spectral response of radiation detectors which have undergone this treatment would, from the amount of trapping and recombination, allow an estimate of the impurity content. It is intended to make these investigations in the future.
A disadvantage of this method is the amount of Ge which is used to make the Ga- In-Ge alloy at the high temperatures. At 850 C the solubility of Ge in the liquid phase with Ga is ~80 atom percent while with In it is ~70 atom percent. After the high temperature cycle it may be necessary to lap off a considerable amount of material. There is considerable evidence also indicating that the Ge lattice beneath this alloy is strained. The resistivity of this strained material is unaltered but the lifetime is decreased. It is probably desirable to lap of this strained material.

This method then is successful in preventing Cu contamination of, and extracting Cu contamination from Ge sample, as far as can be deduced from resistivity measurements. It has a disadvantage in that an appreciable amount of Ge is lost. However it does allow an Sb diffusion without Cu contamination.